REVIEW ARTICLE

# Review on DNA Cryptography

**Mandrita Mondal[1*], Kumar S. Ray[2]**

[1]*Indian Statistical Institute, 203, B.T. Road, Kolkata 700108, India.*
[2]*Department of Computer Engineering and Applications, GLA University, Mathura, Uttar Pradesh, India.*

## Abstract

Cryptography is the science that secures data and communication over the network by applying mathematics and logic to design strong encryption methods. In the modern era of e-business and e-commerce the protection of confidentiality, integrity and availability (CIA triad) of stored information as well as of transmitted data is very crucial. Deoxyribonucleic acid (DNA) is a genetic molecule consisting of two linked strands that wind around each other to form a double helical structure. The backbone of each strand is made of alternating deoxyribose sugar and phosphate groups. To each sugar one of four bases are attached i.e., adenine (A), cytosine (C), guanine (G) and thymine (T). DNA molecules, having the capacity to store, process and transmit information, inspires the idea of DNA cryptography. It is the rapid emerging unconventional techniques which combines the chemical characteristics of biological DNA sequences with classical cryptography to ensure non-vulnerable transmission of data. This innovative method is based on the notion of DNA computing. The methodologies of DNA cryptography are not coded mathematically; thus, it could be too secure to be cracked easily.

## 1. Introduction

In the modern era of e-business and e-commerce "cyber security" is a very crucial term. Itis very important to protect confidentiality, integrity and availability (CIA triad) of stored information as well as of transmitted data. Cryptography is the keystone of modern electronic security technologies. It is the science that secures data and communication over the network by applying mathematics and logic to design strong encryption methods. In other words, cryptography can be defined as the practice and study of techniques to convert original message into human unreadable code.

**Corresponding Author*: *Mandrita Mondal, Indian Statistical Institute, 203, B.T. Road, Kolkata 700108, India; E-mail: mandritamondal@gmail.com*

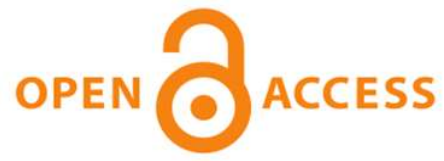

Primarily cryptography was only limited to national defense and diplomacy of government; but, in the modern age cryptography is helpful in every aspect of our daily life, like;providing electronic security to house and offices and business sectors, protecting privacy of ATM, Smart Cards, Credit Card and RFID tags, etc. In the year of 2017, Global State of Information Security® Survey findings reveal that more than 10,000 business and IT executives are adopting innovative cyber-security and privacy safeguards to manage risks and achieve competitive advantages because of the rising threat of hack attacks.

DNA molecules, having the capacity to store, process and transmit information, inspires the idea of DNA cryptography. It is the rapid emerging unconventional techniques which combines the chemical characteristics of biological DNA sequences with classical cryptography to ensure non-vulnerable transmission of data. This innovative method is based on the notion of DNA computing. The methodologies of DNA cryptography are not coded mathematically; thus,it could be too secure to be cracked easily. In the next section a brief overview of modern cryptography is given.

## 2. Cryptography

Cryptology is the study of cryptosystem [1,2]. It isthe science for information security which converts ordinary plain text into human unreadable codes i.e., cipher text and vice versa. Cryptology has two subfields, viz., cryptography and cryptanalysis.

Cryptography is the technique developed by applying mathematics and logic to store and transmit data in coded and secured form so that only the intended recipient can read and process it. The method of securing data by generating cipher text from plain text is also known as encryption. Cryptography protects data from third parties i.e., adversaries and also it is used for user authentication. Cryptanalysis or decryption is the science or technique to decode the cipher text. The basic model of cryptosystem is represented in Figure 1.

The fundamental information security services that cryptography provides are confidentiality, data integrity, authentication and non-repudiation. The cryptographic algorithms, also termed as cryptographic primitives, provide information security as well as cyber security are encryption, hash function, message authentication codes (MAC) and digital signatures. Now we will discuss the basic components of cryptosystem.

- **Plain Text:** It is the original data that is not computationally coded.
- **Cipher Text:** It is encoded or encrypted plain text in the form of human unreadable code.
- **Encryption Algorithm:** It is a mathematical procedure or algorithm that takes plain textas the input and generates cipher text as the output using encryption key.
- **Decryption Algorithm:** It is a mathematical procedure for conversion of cipher text intoplaintext using decryption key. It is the reverse form of an encryption algorithm.
- **Encryption Key**: It is a parameter created explicitly for generating the functional output i.e., cipher text by encryption algorithm.

- **Decryption Key:** It is the parameter which is required to convert the encrypted data i.e.,the ciphertext into its original form i.e., plaintext by the intended receiver.

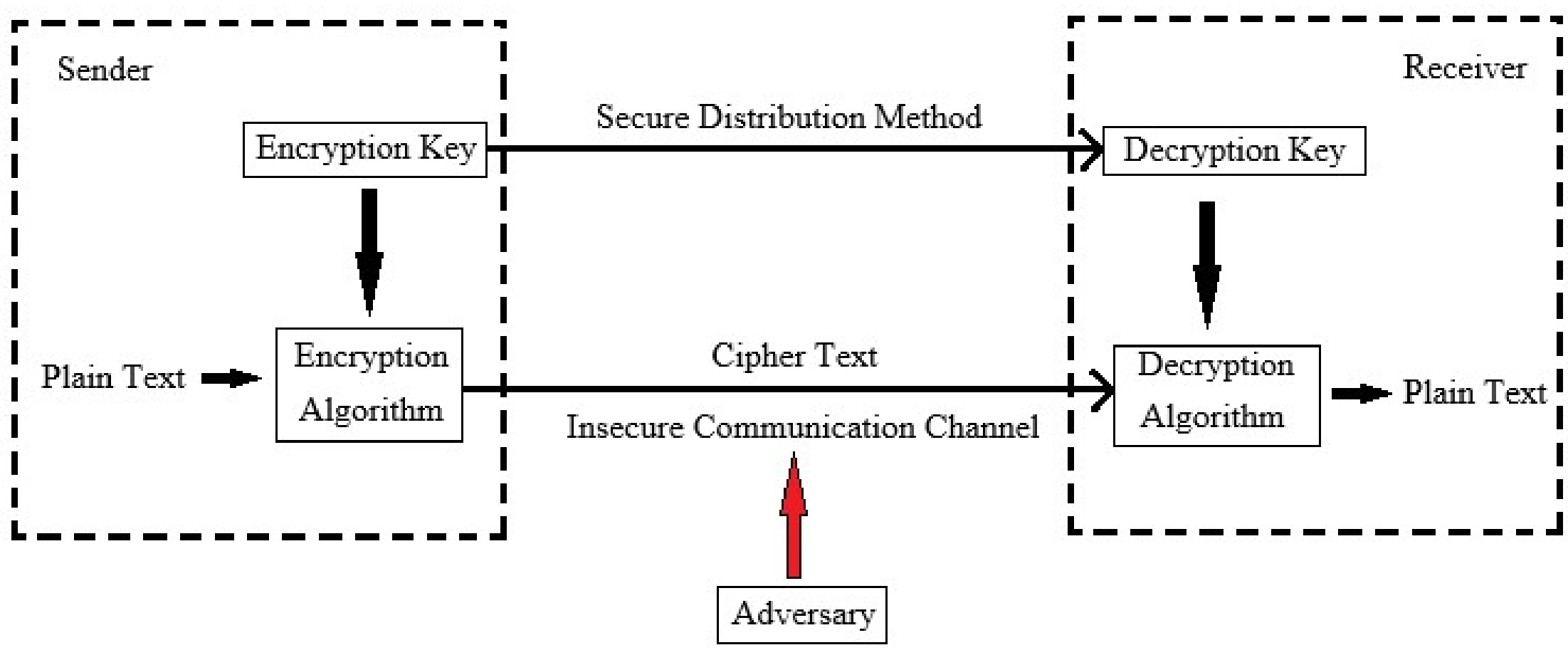

**Figure 1:** *Basic model of cryptosystem.*

Another aspect of cryptography is steganography. It is the practice which not only conceals the content of the secret message but also the fact that the message is being transmitted. Invisiblewatermarking is an example of steganography. In the next subsection we will have a brief discussion on the types of cryptosystems.

## 2.1. Types of cryptosystems

Cryptosystem is of mainly two types based on the techniques of encryption anddecryption.

1. Symmetric key encryption
2. Asymmetric key encryption

### 2.1.1. Symmetric key encryption

Symmetric key encryption or secret key cryptography uses the same key to encrypt and decrypt information. Encryption key is a "shared secret" which is distributed among the senders and receivers. Symmetric encryption is difficult to break for large key size and primarily used forbulk encryption. Confidentiality is the only security service provided by this technique.

Digital Encryption Standard (DES), Triple-DES (3DES), BLOWFISH and IDEA are a few examples of popular symmetric key encryption methodologies. There are certain advantages as well as disadvantages of using this technique in modern cryptography.

**Advantages**

- Symmetric key encryption is faster compared to other methods because of thesmaller key length.
- As no key is transmitted along with the coded data, the possibility of data being intercepted is minimal.

- In symmetric key encryption password authentication is required to verify the recipient's identity.
- Data can be transmitted to a group of recipients who possesses the secret key for decryption which is shared prior to the exchange of original information.

**Disadvantages**

- Key transportation is the huge disadvantage of symmetric key encryption. The secret key is supposed to be shared prior to the exchange of original data. As the electronic communication channels are highly insecure, thus the secret keys are recommended to exchange personally.
- To prevent hack attacks the keys, need to be changed regularly which makes the whole methodology cumbersome and expensive as the number of keys required in symmetric cryptography is huge. The number of keys required for a group of completely connected n participants using symmetric cryptography is: $\frac{n\times(n-1)}{2}$.
- Digital signatures cannot be provided by symmetric cryptography which is non-repudiated.

### 2.1.2. Asymmetric key encryption

Asymmetric key encryption or public key cryptography uses different keys to encrypt and decrypt information. In this methodology, each of the participants of the communication has two keys; one is a public key which is shared with all the participants, and the other is private key which is secret and only the intended receiver knows it. Though the public and private keys are apparently different, these are mathematically related. Each of the public keys has a corresponding private key. This technique can provide integrity, authentication, and non-repudiation.

RSA (Rivest–Shamir–Adleman), DIFFLE, Elliptic Curve cryptography are a few popular examples of asymmetric key encryption algorithms. Now we will give a brief discussion on the advantages and disadvantages of using asymmetric cryptography.

**Advantages**

- In asymmetric encryption method the key transportation problem is eliminated.
- Users can be added to and removed easily from asymmetric cryptosystem. The same pair of public and private keys is required to communicate with all the participants of asymmetric cryptosystem. Thus, the addition of a new user needs the generation of only one public key-private key pair. The key revocation mechanism effectively cancels a key when the user is removed.
- Asymmetric key encryption methodologies are more secure than symmetric key encryption, as private keys are not required to be transmitted to anyone.
- Digital signatures can be provided by asymmetric cryptography which is repudiated.

**Disadvantages**

- Asymmetric cryptography is significantly slower than symmetric key encryption because of the length of keys. For secure transmission of large data, users generally use asymmetric key encryption techniques to establish a connection and then share symmetric secret key and uses symmetric cryptography.

After giving a brief overview on the types of cryptosystems, we will now discuss the attacks on cryptosystem.

## 2.2. Attacks on cryptosystem

Attacks on cryptosystem can be defined as an intelligent, deliberate act to break the security services illegally. Attack of the adversary can be either passive or active. Passive attack is the unauthorized access of information without affecting the system resources. In active attack, the adversary attempts to affect system resources and alter their operation. Depending onthe attacking techniques the attacks can be categorized as follows.

1. **Ciphertext-only attack (COA):** In this case, the attacker has access only to the set of ciphertext and tries to extract the plaintext or the key.

2. **Known plaintext attack (KPA):** In KPA the attacker has access to the plaintext (i.e., crib) and the corresponding encrypted ciphertext. The attacker tries to deduce secret keys and code books.

3. **Chosen Plaintext attack (CPA):** Here, the attacker has the capability to choose random plaintext and has an on the corresponding encrypted ciphertext. The attacker tries to reduce the security of the cryptosystem by revealing further secret information.

4. **Chosen Ciphertext attack (CCA):** In this case, the attacker has the ability to choose a ciphertext and decrypt it to study the corresponding plaintext.

So far, we have discussed the terminology, categorization of cryptography and the attacks on the cryptosystem. In the next section we will focus on different modern methodologies ofcryptography.

## 2.3. Methodologies of cryptography

Before the modern era, cryptography was only used to keep message confidentiality by the military leaders, spies, and diplomats. By the passing years encryption techniques have been evaluated to ensure secure computation, to check message integrity, to authenticate senders' and recipients' identities etc.

Cryptography was originated in Egypt around 4000 years ago, in 1900 B.C. Since then, cryptography has gone through massive evolution. In ancient Egypt hieroglyphics were drawn ontombs, which are actually substitution cipher, where one character or symbol was substituted for another. Another model of ancient cryptosystem is Scytale which was developed in Greece for sharing secret information in about 500 B.C. This device used to work following the methodologies of transposition cipher in which the order of the characters was changed.

Cryptography was first used for military purposes nearly 2,000 years ago when Julius Caesar developed monoalphabetic substitution cipher method to send confidential information during war. This earliest recorded Roman encryption method, known as the Caesar Shift Cipher (Figure 2), shifts the alphabets of the original text by a predetermined number (cipher key). Onlythe intended recipient can decode the encrypted text.

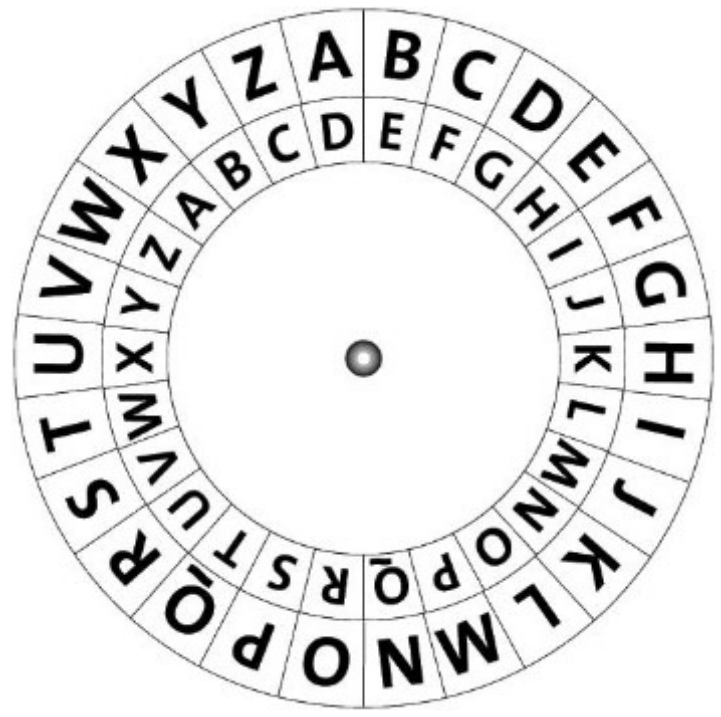

**Figure 2:** *Caesar shift cipher.*

In 15th century Blaise De Vigenere developed Vigenere Cipher. This improved encryption technique follows the method of polyalphabetic substitution cipher method where the key is changed throughout the encryption process.

During the First World War, Zimmerman Telegram and Choctaw Codetalkers are significant examples of cryptography. In the Second World War an electro-mechanical encryption machine Enigma by Germans became popular. Another efficient encryption device called Purple was discovered by Japan during the war period.

Modern cryptography follows strong scientific methods to design encryption algorithms which are theoretically unbreakable by an adversary. But in practical, computationally secure mechanisms is sometimes less feasible to do so. Now, we are going to give an outline of the present state of the art of cryptography.

### 2.3.1. One-time pad (OTP)

One-time pad algorithm was developed towards the end of 19th century from Vernam Cipher. If the key used in OTP is randomly generated and not used more than once, then the algorithm is considered to be completely unbreakable. Previously, the randomly generated keys were shared as a pad of paper, so the sheets could be torn off after the use of the key; thus, the algorithm was termed as one-time pad.

The secret key of the algorithm is generally a string of characters or numbers which is at least as long as the longest message to be encrypted. One time pad is a binary additive stream cipher, where one bit of plain text is encrypted at time by an 'exclusive OR' (XOR) addition withthe corresponding bit in the secret key. The keys used for encryption and decryption are the same. This algorithm is explained using an example in Figure 3.

Pseudo-Random Number Generators (PRNGs) are algorithms which use mathematical formulas to generate sequences of random numbers. Some popular PRNGs are lagged Fibonacci generators, linear congruential generators, linear feedback shift registers etc. For cryptographic applications Cryptographically Secure Pseudo-Random Number Generators (CSPRNGs) are used and it is more protected than PRNGs. CSPRNG requires passing the

next- bit test i.e. if the first *k* bits of a random sequence is given, there is no algorithm for prediction of the ($k$+1)th bit with probability of success non-negligibly better than 50% within polynomial time [2]. The CSPRNGs which can pass next-bit test, is capable of passing all other polynomial-time statistical randomness test [3] and can withstand massive hack attacks.

**ENCRYPTION**

| | |
|---|---|
| **Message to be Encoded** | DNA |
| **Plain Text (Corresponding Bit String)** | 01000100 01001110 01000001 |
| **Randomly generated Key** | 11010001 01101000 00101011 |
| **Cipher Text (XOR between plain text and key)** | 10010101 00100110 01101010 |

**DECRYPTION**

| | |
|---|---|
| **Cipher Text** | 10010101 00100110 01101010 |
| **Key** | 11010001 01101000 00101011 |
| **Plain Text (XOR between cipher text and key)** | 01000100 01001110 01000001 |
| **Decoded Message** | DNA |

**Figure 3:** *Example of one-time pad method.*

### 2.3.2. Symmetric key cryptography

Symmetric key cryptography or secret key cryptography was primarily developed for bulk encryption of data or stream of data. It uses the same key for encryption and decryption of secret information. The block cipher is the symmetric encryption algorithm that encrypts fixed-length groups of bits i.e., block of the plain text at a time. On the other hand, another type of symmetric encryption algorithm, stream cipher, can encrypt a single bit or byte of data at a time. We have discussed the advantages and disadvantages of symmetric key encryption in subsection 2.1.

Symmetric encryption algorithms used now a day are generally block ciphers. The usual sizes of the blocks which the block cipher can encode at a time are 64 bits, 128 bits and 256 bits. Generally, a 128-bit block cipher encrypts 128 bits of plain text at a time and generates cipher text of 128 bits. The most popular block ciphers having a wide range of applications are DES, Triple DES, AES, Blowfish, Twofish etc.

- ***DES -*** Data Encryption Standard (DES) is a popular 64-bit block cipher developed in 1975 and standardized by ANSI (American National Standards Institute) in 1981 as ANSI X.3.92. DES is supposed to be vulnerable to brute force attacks. Though, the key used in DES consists of 64 bits, the effective key length is 56 bits. The 8 bits of actual key size are used for checking parity and then discarded.

- ***Triple DES -*** The basic methodology of Triple DES is based on DES. It encrypts theplain text three times. The overall key length of Triple DES is 192 bits i.e., three 64-bit keys are used in this block cipher. Like DES, the effective length of each key is 56 bits.In Triple DES, the first encrypted cipher text is again encrypted by the second key and again the third key encrypts the resulting encrypted cipher text. Though this algorithm is much more secure than DES, but it is too slow for many real-life applications.

- ***AES -*** The most commercially popular symmetric block cipher is Advanced Encryption Standard (AES) which has a wide range of applications in modern security system suchas financial transactions, e-business, wireless communication, encrypted data storage etc. It is more secure and faster than triple DES both in hardware and software. The numberof rounds as well as the length of the keys of AES is variable. There are 10 rounds for 128-bit key, 12 rounds for 192-bit key and 14 rounds for 256-bit key. Different 128-bit round keys, calculated from original AES key, are used for each round. AES algorithm treats the block of data in bytes, i.e. 128 bits of plain text is treated as 16 bytes. The development of AES, a transparent successor to DES, was started by National Institute ofStandards and Technology (NIST) in 1997 and is more secure against brute forceattacks.

- ***Blowfish –*** Blowfish is another 64-bits block cipher which supports different key lengths, varies from 32 to 448 bits. This feature makes the Blowfish algorithm ideal for domestic as well as exportable use. This algorithm can be efficiently used in software and is free for all users as Blowfish is unpatented and license free.

- ***Twofish -*** It is 128-bit symmetric key block cipher and supports the key length up to 256 bits. Twofish is related to above mentioned block cipher Blowfish but is not as popularas Blowfish.

Stream ciphers are commercially less popular than block ciphers because of its difficulty in implementation. Besides stream ciphers do not ensure integrity protection or authentication. These algorithms are useful when the amount of data is either unknown or continuous, for example network streams.

Rivest Cipher 4 (RC4), also known as ARCFOUR or ARC4, is the most popular algorithm among stream ciphers. It has a wide range of applications in software system; for example, SSL/TLS, Microsoft Lotus, Oracle Secure SQL, Microsoft Windows and Apple OCE etc. RC4 has been applied in protocols like WEP (Wired Equivalent Privacy) and WPA (Wi-Fi Protected Access). Now the usage of this algorithm is descending because of the recent revelation of vulnerabilities in RC4.

Other stream ciphers which are not as popular as RC4 are A5/1, A5/2, FISH, Chameleon, Helix, Panama, SOBER, WAKE etc.

### 2.3.3. Asymmetric key cryptography

Asymmetric key cryptography or public key cryptography utilizes different but mathematically related keys; public keys for encryption which are generally distributed widely, and private keys for decryption which are known only to the intended recipients. In subsection 2.1 we have discussed the advantages of asymmetric key cryptography over symmetric key cryptography as well as the disadvantages of this cryptosystem.

In 1976, Whitfield Diffie and Martin Hellman projected Diffie-Hellman key exchange protocol [4] which is a revolutionary work in the history of cryptography. In the year 1978 Ronald

Rivest, Adi Shamir, and Len Adleman proposed RSA algorithm [5], another landmark of asymmetric key cryptography. Some other methodologies that public-key encryption includes are Cramer–Shoup cryptosystem, ElGamal encryption and different elliptic curve techniques. Now we will focus on the popular algorithms of asymmetric key cryptography.

**Diffie-Hellman Key Exchange Protocol**

In 1974, Malcolm J. Williamson first developed the protocol [6] which isnow popular as Diffie-Hellman key exchange [4]. As he was then working at the Government Communications Headquarters (GCHQ), Britain, was therefore unable to publish his research work. Two years later, Whitfield Diffie and Martin Hellman published the paper on Diffie-Hellman key exchange or exponential key exchange protocolwhich is scheme to exchange information over a vulnerable public channel.

Let two people Alice and Bob, as referred to in cryptographic literature, want to share data securely through public channel. The core idea to build a secure communication platform by this protocol is that there is some secret information known only to Alice and Bob. This information is used to derive a suitable key by which the data can be encoded or decoded. The algorithm of Diffie-Hellman key exchange protocol (Figure 4) is as follows.

- ***Step 1:*** Alice and Bob be in agreement over a large prime number $p$ and a base $g$; where, $1< g < p$.
- ***Step 2:*** Alice selects a secret number $a$, and sends the value of ($g^a$ mod $p$) to Bob.
- ***Step 3:*** Bob selects a secret number $b$, and sends the value of ($g^b$ mod $p$) to Alice.
- ***Step 4:*** Alice calculates the value of (($g^b$ mod $p$)$^a$ mod $p$).
- ***Step 5:*** Bob calculates the value of (($g^a$ mod $p$)$^b$ mod $p$).
- ***Step 6:*** The numbers calculated by Alice and Bob in step 4 and 5 are the same number which is used as the key.

Thus, they both have the key without disclosing any secret numbers to each other. The key is then used to communicate securely by a chosen cryptosystem. In this protocol $p$ and $g$ need not to be protected. The security of widely used Diffie-Hellman protocol is based on the computational number theory problem called the discrete logarithm problem which is too hard tosolve in polynomial time.

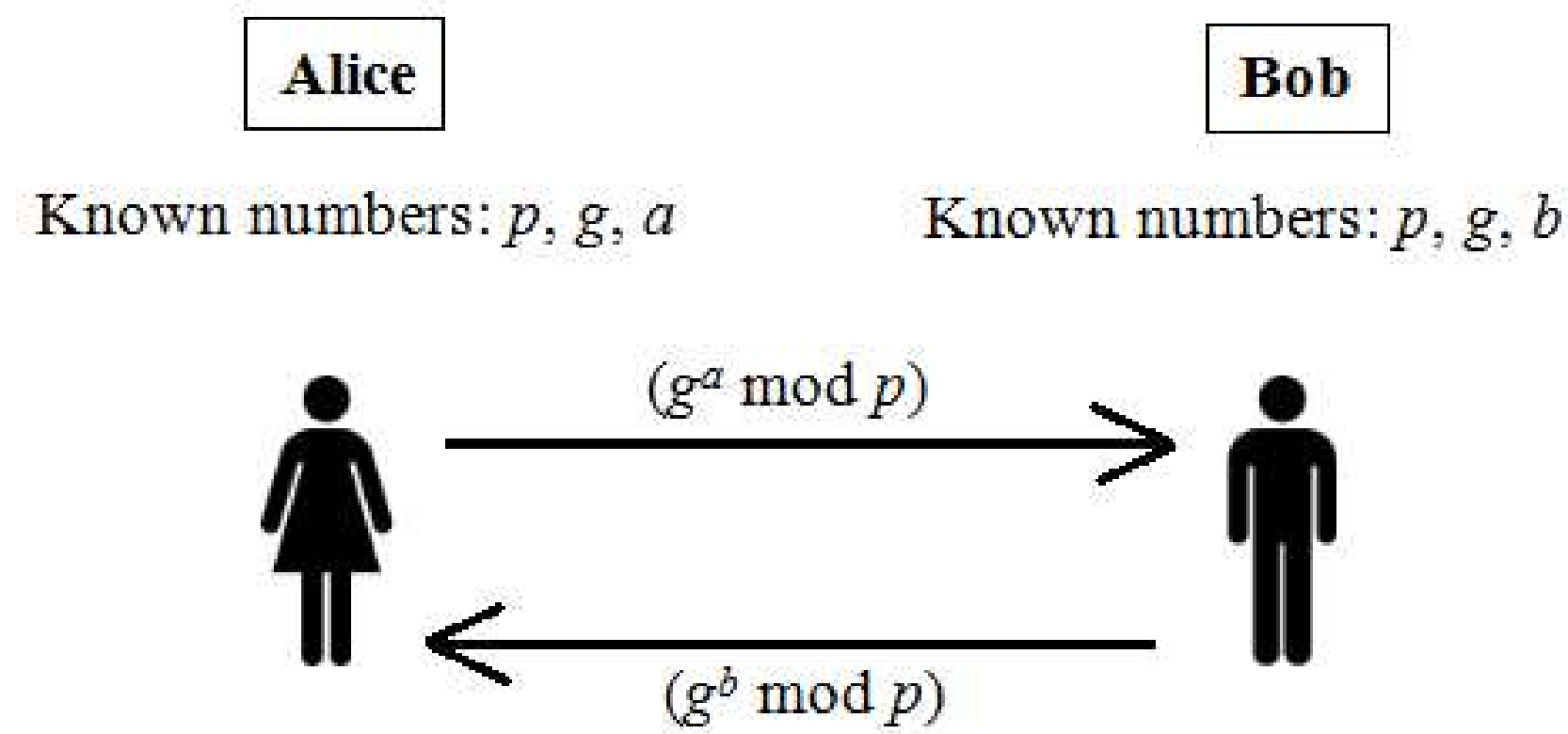

**Figure 4:** *Pictorial representation of Diffie-Hellman key exchange protocol.*

**RSA Algorithm**

RSA algorithm [5], the most widely used asymmetric cryptosystem, was presented by Ron **R**ivest, Adi **S**hamir and Leonard **A**dleman of the Massachusetts Institute of Technology in 1977. The security of RSA algorithm depends on the fact that the factorization of a large number, which is a product of two large prime numbers, is too difficult to solve. It is a trapdoor function which is easy to compute in one direction but difficult to compute the reverse without specific information. For example, let, $P \times Q = N$, where P and Q are very large prime numbers. If $P$ and $Q$ are given, N can be computed very easily. But if $N$ is given, the factorization of $N$ cannot be computed in polynomial time. But, if the value of $Q$ is known along with $N$, then $P$ can be computed easily i.e., $P$ = Q.

There are two aspects of RSA algorithm; first is the generation of the pair of keys, one is public key and the other is private key; second is the encryption and decryption methodology. The steps for generation of keys are stated below.

**Step 1:** Generation of the RSA modulus ($n$)

- Two large prime numbers p and q are selected.
- The value if n is calculated; where $n = p \times q$. To make the encryption technique more secure, n should be minimum of 512 bits i.e., very large number.

**Step 2:** Derivation of number ($e$)

- The number e must be $1 < e < (p - 1)(q - 1)$.
- $e$ and $(p - 1)(q - 1)$ should be co-prime.

**Step 3:** Formation of the public key

- The RSA public key is formed by the pair of numbers (n, e) and published publicly.
- As the public key is formed using $n$, thus, it will be difficult for the invader to derive the two prime numbers p and q from n in polynomial time to break the encrypted data. This trick increases the security of RSA algorithm.

**Step 4:** Generation of the private key

- The unique private key d is derived from $p$, $q$ and $e$. This key is kept secret.
- The number d is derived from the equation, $d = e{-1} \ mod \ (p - 1)(q - 1)$. If the values of $p$, $q$ and $e$ are given, then Extended Euclidean Algorithm derives the value of $d$ as the output.

The next aspect of RSA algorithm is encryption and decryption of the data. After generation of the keys, this process is relatively uncomplicated.

**Step 1: Encryption methodology of RSA**

- RSA operates on the numbers modulo $n$, thus, the plain text should be converted into a sequence of numbers less than n before encryption process. Here, the public key to encrypt the plaintext is ($n$, $e$).

- If a plaintext is represented by $P$ and the corresponding ciphertext is represented by $C$, then, the equation to derive $C$ is represented by Equation 1.

$$C = Pe \ mod \ n \qquad [1]$$

Here, the value of $C$ is also less than $n$.

**Step 2: Decryption methodology of RSA**

- The intended recipient who has the private key d, decrypt the ciphertext C using the Equation 2.

$$P = Cd \ mod \ n \qquad [2]$$

To increase the security of RSA encryption the chosen prime numbers $p$ and $q$ should be large enough and $e$ should not be a very small number. Otherwise, the encryption and decryption would be non-one-way functions and the algorithm would become breakable.

**ElGamal Cryptosystem**

ElGamal cryptosystem [7] is an asymmetric key encryption which was proposed by Tather Elgamal in 1984. The security of this algorithm is based on the Discrete Logarithm Problem. For a given number, there is no existing algorithm which can find its discrete logarithm in polynomial time, but the inverse operation of the power can be derived efficiently. Another key aspect of ElGamal cryptosystem is randomized encryption. This algorithm can establish a secure channel for key sharing and is generally used as key authentication protocol.

For security the key size of this algorithm should be greater than 1024 bits. The major drawback of the ElGamal algorithm is that it is relatively time-consuming.

**Elliptic Curve Cryptography (ECC)**

Elliptic-curve cryptography (ECC) [8] the promising future of asymmetric key encryption, is based on the algebraic structure of elliptic curves over finite fields. Similar to ElGamal cryptosystem, the security of ECC also depends on the algorithmically hard discrete logarithm problem. Though this algorithm follows the same kind of technique as Diffie-Hellman key exchange protocol and RSA algorithm, but the unique feature of Elliptic-curve cryptography is that the numbers are chosen to form a finite field defined within an elliptic curve expression. An example of an elliptic curve graph is presented in Figure 5. The key size of this technique is much smaller; for example, ECC with 160 bits key length ensures same level of security as RSA algorithm with key length of 1024 bits. As the processing power consumption and required memory sizes are significantly low, thus, there is a huge possibility of applications of this relatively new algorithm on constrained devices. ECC has several applications in modern life activities: for example, secure data transmission, digital signature, mutual authentication etc.

### 2.3.4. Steganography

The word 'steganography' comes from the Greek words 'stegano' which means 'covered' and 'graphie' means 'writing'. Instead of performing literal encryption, this methodology ensures data security through obscurity. It transmits data by embedding it in an unnoticeable way, through video, audio, document or image files. In 18th century during the Revolutionary War, both the British and American armies used invisible ink which is a mixtureof ferrous sulfate and water. Experts claim that Leonardo Da Vinci embedded hidden messages in his painting. These are examples of ancient practices of steganography.

Now, there are several techniques to conceal data in normal files. One of the most popular spatial domain image steganography methodologies is least significant bit technique or *LSB* [9]. In a digital system, colors are represented as additive combinations ofprimary colours i.e., red (R), blue (B) and green (G), whose values are assigned from minimum 0 to maximum 255 in decimal and from 00000000 to 11111111 in binary. To encrypt the message the last few insignificant bits in a byte of an image is changed in LSB technique. This makes an unnoticeable change in colour which is nearly impossible to detect in naked eyes, thus, encrypteddata can be securely transmitted. To make LSB more robust, basic LSB technique has been modified in several ways. But, for ASCII text this methodology does not work as the modification of single bit entirely changes the character. Other techniques of image steganography are Transform Domain technique, Discrete Cosine Transform Coefficient technique etc. LSB-based Audio Steganography and Audio wave steganography are a few techniques for audio steganography. LSB insertion method on video images is a popular method for video steganography.

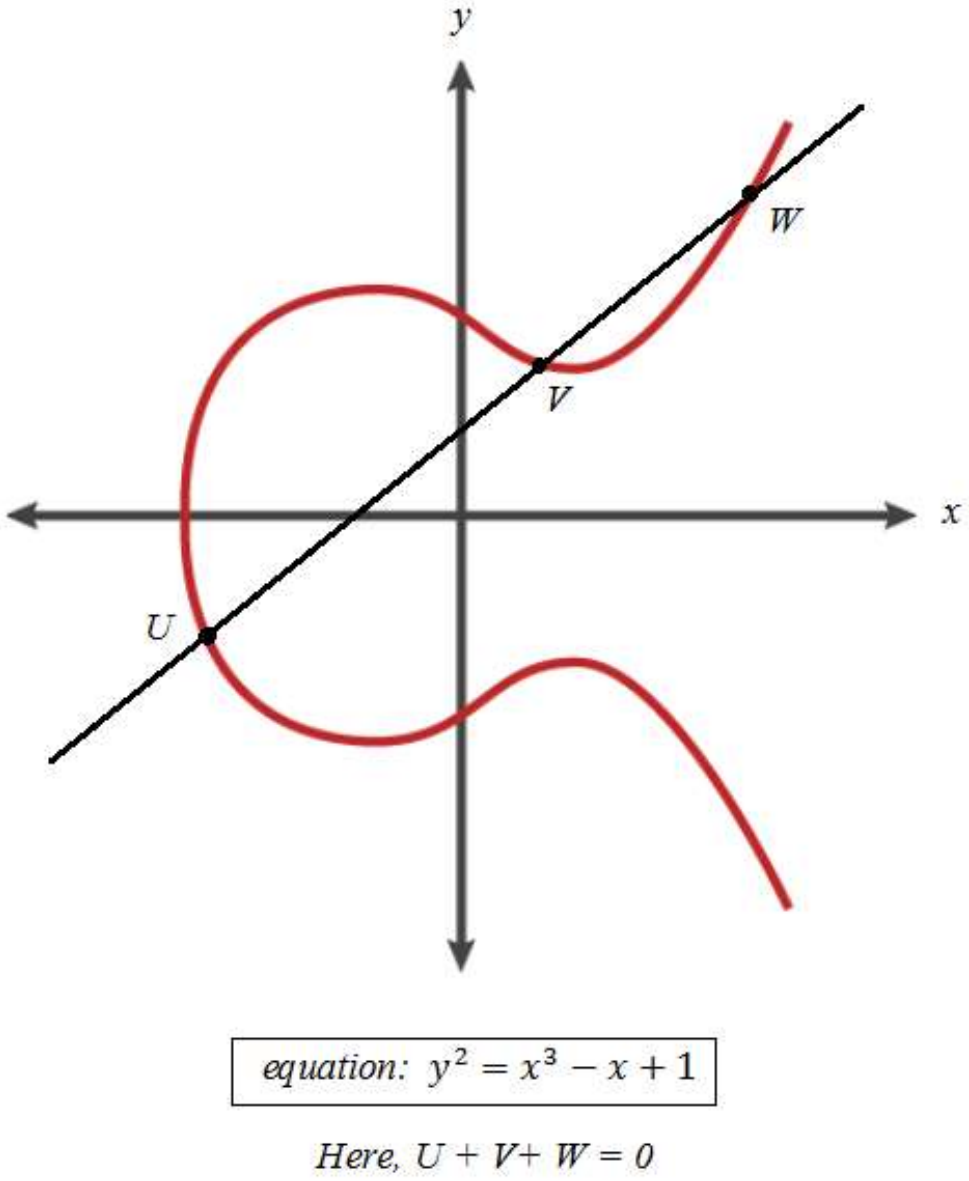

**Figure 5:** *Elliptic-curve graph.*

## 2.4. Unconventional techniques of cryptography

The amount of electronically stored and communicated information is increasing dramatically day by day. More protection and control over the information assets is needed as the escalating tech literacy is a massive threat of the modern era. As a result, there is a huge requirement for a more secure and unbreakable cryptosystem. The security of widely applied modern encryption algorithms are generally based on the trapdoor functions, but, after the invention of DNA computing and quantum computing, this function can be solved in polynomial time. These natural and unconventional computations are becoming practical reality and thus the existing encryption methodologies are becoming vulnerable. In this section we will have a brief discussion on upcoming methods of cryptography which will ensure more indissoluble and secure data transmission as well as data storage.

### 2.4.1. Quantum cryptography

Quantum cryptography is the flourishing encryption technique based on the properties of quantum mechanics, basically Heisenberg Uncertainty principle and the principle of photon polarization. This methodology exploits the counterintuitive behavior of elementary particles at atomic scale, generally the photon particle. In quantum cryptography information is transmitted by quantum bit, also called qubit, which is actually a single photon particle. The application of quantum information theory was first proposed by Stephen Wiesner in the late sixties [10]. The real-life implementation of quantum cryptography was established by the path breaking paper of Charles Bennett and Gilles Brassard in 1984 [11]. A protocol for quantum key distribution (QKD), also called BB84 protocol, was proposed in this paper. Today, big tech giants such as IBM and Google are investing a lot in quantum AI laboratories and trying to commercialize quantum computing as well as quantum cryptography.

### 2.4.2. DNA cryptography

Another rapidly emerging methodology in the domain of cryptography is based on DNA sequences. DNA molecules having the capacity to store, process and transmit information inspiresthe idea of DNA cryptography. It works on the concept of DNA computing which uses fourbases i.e., Adenine (A), Guanine (G), Cytosine (C) and Thymine (T) to perform computation. One of the advantages of DNA computation is the massive parallelism of DNA molecules. In an *in vitro* assay about $10^{18}$ processors working in parallel can be easily handled. Because of this huge parallelism the trapdoor function, which is the basic security secret of most of the traditional cryptosystems, can be solved in polynomial time. Thus, it is high time to find the alternative of traditional cryptosystem. As the mathematical aspect of cryptography is being replaced by DNA chemistry in the domain of DNA cryptography, thus this technique is virtually unhackable by conventional methods as well as quantum computation.

Today, different research works are being performed across the globe either to enhance the existing DNA cryptography methodologies or to propose innovative and novel approaches in this domain. In the coming ten to thirty years when DNA computers will be commercially available, it will take over modern silicon-based technology. Microsoft has taken the initiative to explore thedomain of DNA computing where Luca Cardelli is the leading name in this field. In the next section we will discuss innovative proposals in the field of DNA cryptography.

## 3. DNA Cryptography

Besides the huge parallelism, DNA molecules also have massive storage capacity. A gram of DNA molecules consists of 1021 DNA bases which is nearly about 108 tera-byte. Thus, it can be concluded that a few grams of DNA can restrain all the data stored in the world. These advantages of DNA computation motivate the idea of DNA cryptography. The pioneer of DNA computing is Prof. Leonard Adleman who is also known as the 'A' of RSA algorithm (subsection 2.3.3). Though DNA cryptography is in the early stage of maturation, but in last few years several methodologies for designing cryptosystem and steganography have been proposed.

### 3.1. DNA Substitution and one-time pad [12]

Gehani et al. [12] first proposed the possibility of designing cryptosystem using DNA molecules in 1999. They developed one-time pad encryption methodology using two different techniques; the first one is by DNA substitution method using libraries of randomly generated distinct pads represented by DNA strands and the second one is by using bit-wise XOR scheme using molecular computation.

#### 3.1.1. One-time pad scheme by DNA substitution

We have explained one-time pad scheme of classical cryptography, which is supposed to resist all kind of hack attacks, in the subsection 2.3.1. The cryptosystem designed by [12] using one-time pad methodology contains the following:

1. Plaintext binary message which is represented by DNA strands of length n and spitted up into words of fixed length.

2. Codebook library containing huge number of long DNA pads; each representing unique and random mapping of plaintext word to cipher word. It serves as the key of the proposed scheme. The one-time pad codebook consists of the repeating unit (i) which has three domains (Figure 6).
    - The first domain $C_i$ represents the set of words.
    - The next domain $P_i$ represents the corresponding plaintext words.
    - The last domain is the stopper sequence which acts as punctuation between the repeating units.

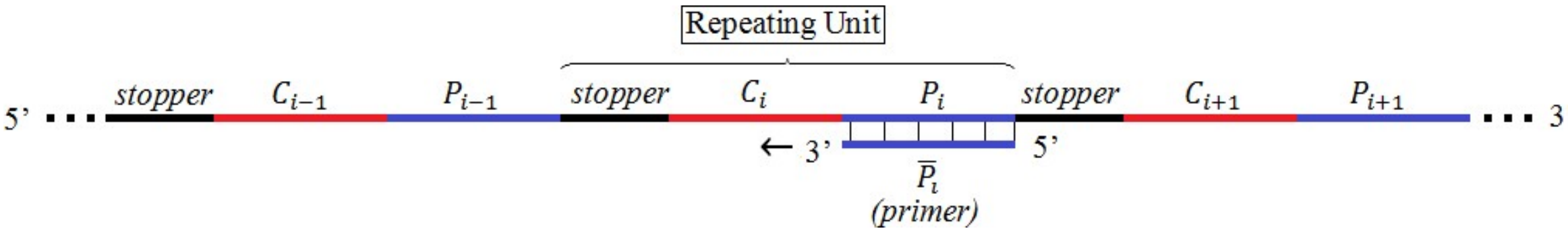

**Figure 6:** *Repeating unit of DNA codebook.*

The one-time pads are formed secretly and shared between the sender and the receiver before transmission of the message. The synthetic short oligonucleotides representing the repeating unit are randomly assembled, then isolated and cloned to form huge one-time pad codebook. Commercially available DNA microarray or DNA chip, which is a collection of microscopic DNA spots attached to a solid surface, can also be used as DNA codebook. According to the DNA codebook, each plaintext word is substituted by the corresponding cipher word. The domain $P_i$ of the repeating unit is the site where the oligonucleotide, $\overline{P_i}$, hybridizes. Here, $\overline{P_i}$, which is complementary to $P_i$, acts as a primer. It is then elongated using the polymerase enzyme following the protocol of DNA replication. The stopper sequence prevents the further extension of the newly synthesized strands. As a result, a unique set of plaintext-ciphertext word pair is formed. Then word-pairs are cleaved, and the plaintext domain is removed. Thus, the input message is encrypted in the form of DNA cipher word. Decryption is the reverse operation of encryption.

### 3.1.2. One-time pad scheme by bit-wise XOR

Another one-time pad technique proposed by [12] is based on the fundamental principle of Vernam cipher [13]. The algorithm is stated below.

**Step 1:** The key sequence *S* (one-time pad), which contains random *R* number of bits, is distributed to both sender and receiver in advance. These bits are used for encryption of the plaintext. The variable *L* represents the number of unused bits in *S*. Initially, *L* = *R*.

**Step 2:** The plaintext is represented by *P* which contains *N* bits such that *N* < *L*. Each bit $P_i$ (where, *i* = *1,……, N*) is XOR'ed with $K_i = S_{R-L+i}$ to generate the encrypted cipher bit $C_i$, where $C_i = P_i \oplus K_i$. The truth table of XOR is given below (Table 1).

**Table 1:** *Truth table XOR.*

| $P_{\mathrm{i}}$ | $K_{\mathrm{i}}$ | $C_{\mathrm{i}}$ |
|---|---|---|
| 0 | 0 | 0 |
| 0 | 1 | 1 |
| 1 | 0 | 1 |
| 1 | 1 | 0 |

**Step 3:** The used bits are destroyed from the source pad sequence *S*.

**Step 4:** The cipher sequence $C = C_1, C_2, \dots \dots, C_N$ is generated after encrypting all the bits of *P*.

**Step 5***:* To decrypt the cipher text the receiver again performs XOR using each bit of cipher sequence ($C_i$) and the pad sequence ($K_i$ where, $K_i = S_{R-L+i}$) to generate the plaintext bit $P_i$ i.e. $C_i \oplus K_i = P_i$.

This algorithm is implemented in DNA cryptography using the self-assembly of DNA tiling [14-17]. The TAO triple-helix tiles used in [12], pictorially represented in Figure 7.

Four hybridized single stranded DNA sequences form TAO triple-helix tile as shown in Figure 7. The strands of upper and lower helices have bare ends, but the ends of the central helix are capped with hairpin. The tiles are capable of self-assembly, and a huge complex structure can be generated by hybridization of the sticky ends of the tile to the neighbour tiles [12].

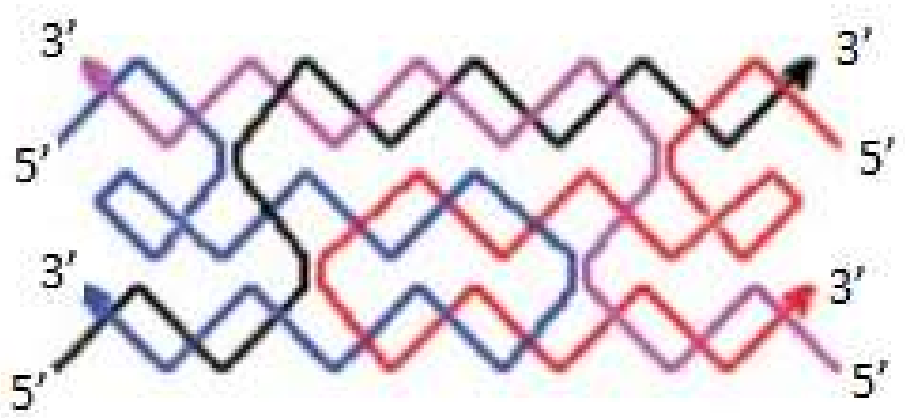

**Figure 7:** *TAO triple-helix tile.*

The Vernam cipher is implemented in DNA cryptography using the following steps:

**Step 1:** The $n$ bits of the plaintext are encoded in the form of DNA sequences and using appropriate linking sequences the scaffold strand $a_1 a_2 \ldots \ldots a_n$ is generated.

**Step 2:** The further portion of the scaffold strand $a'_1 a'_2 \ldots \ldots a'_n$ can be formed using arbitrary inputs which is actually the one-time pad of the scheme.

**Step 3:** The input assembly (Figure 8) i.e., the tile structure is formed by using the two types of input scaffold strands (mentioned in step 1 and step 2), suitable linking sequences and different other sequences [18].

**Step 4:** An opening for hybridization of a single output tile is created by the input assembly because of the un-complemented sticky ends. After the introduction of the output tiles (i.e., encoded ciphertext) because of the self-assembly of DNA tiles, it can attach to the desired sticky ends.

**Step 5:** Reporter strand $R$ (Equation 3) is formed by treating the assembly with DNA ligase.

$$R = a_1 a_2 \ldots \ldots a_n . a'_1 a'_2 \ldots \ldots a'_n . b_1 b_2 \ldots \ldots b_n \text{ where, } b_i = a_i \oplus a'_i \quad [3]$$

**Step 6:** The reporter strand is denatured to form single strands and made run through polyacrylamide gel. The strand can be extracted from the gel.

**Step 7:** The reporter strand contains three domains. The first domain $(a_1 a_2 \ldots \ldots a_n)$ encodes the input plaintext, the second domain $(a'_1 a'_2 \ldots \ldots a'_n)$ encodes one-time pad i.e. the key and the third domain $(b_1 b_2 \ldots \ldots b_n)$ represents the ciphertext. If a restriction site is encoded between the second and third domain, the ciphertext can be recovered from the reporter sequence by treating it with the corresponding restriction endonuclease.

**Step 8:** The gel purification is performed with the sequence encoding the ciphertext and it is sent to the receiver.

**Step 9:** According to the principle of the Vernam cipher, the same key sequence $(a'_1 a'_2 \ldots \ldots a'_n)$ is used for decryption. The plaintext can be recovered by using scaffold strands representing ciphertext $(b_1 b_2 \ldots \ldots b_n)$ and key sequence as input assembly.
The intended output is, $a_i = b_i \oplus a'_i$.

**Step 10:** After self-assembly of DNA tiling, the reporter strand $R'$ (Equation 4) is recovered following the step 6.

$$R' = b_1 b_2 \ldots \ldots b_n . a'_1 a'_2 \ldots \ldots a'_n . a_1 a_2 \ldots \ldots a_n \quad [4]$$

**Step 11:** Finally, the third domain of $R'$ representing the plaintext is recovered following previously explained step 7 and 8.

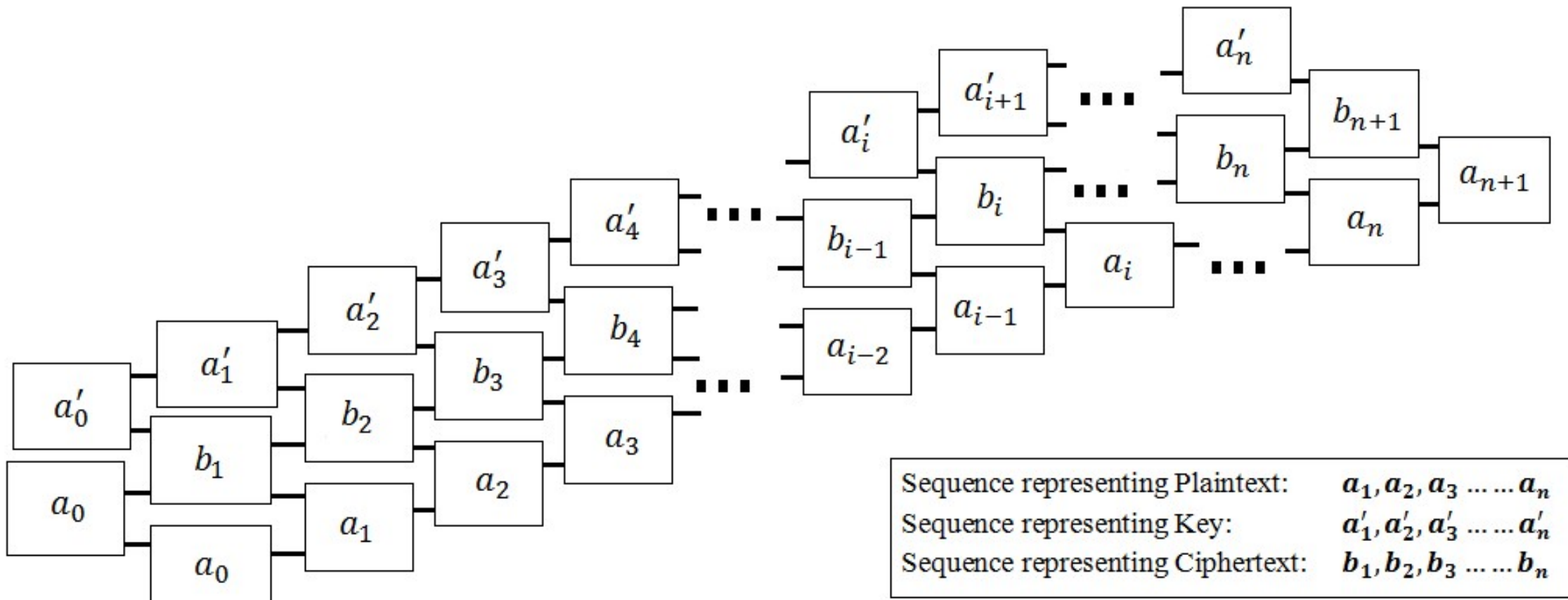

**Figure 8:** *Representation of XOR operation using DNA tiles.*

### 3.1.3. DNA steganography

The research work of [12] also proposed a method of steganography using DNA sequence..

**Step 1:** The plaintext is represented in the form of input DNA strands.

**Step 2:** The input strands are tagged with secret keys which are also in the form of DNA sequence.

**Step 3:** These strands are mixed with random DNA strands which are specified as distracters.

**Step 4:** If the secret key is known to the receiver, then the strands representing the plaintext canbe extracted from the mixed-up solution following affinity purification protocol. The single stranded sequence used in the experiment is the complementary sequence of the secret key.

The disadvantage of this method is that the sequence representing the original text can be recovered based on the entropy difference between the distracter and the input DNA strands i.e., plaintext. This difference can be reduced by designing the distracter sequence similar to the inputDNA sequence. Otherwise, to match the distribution between the distracter and the sequence representing the plaintext DNA-SIEVE [12] can be used. It shapes the set of random distracters into one for distribution match. Another technique of entropy reduction is, if making the plaintext to imitate the distracter. This can be achieved by compressing the plaintext with lossless algorithm [19].

## 3.2. Bi-Layer Steganography using DNA microdots [20]

The bi-layer steganographic technique for transmission of secret message using DNA microdots was demonstrated in [20]. DNA microdots, used for hiding confidential messages, are microscopic DNA spots attached to a solid surface. In this two layered steganographic technique, the original message is first encrypted within the massive human genomic DNA which contains about $3\times 10^9$ base pairs. Then, it is further hidden into microdots. The algorithm of the steganographic procedure is discussed through the following steps.

**Step 1:** The original text is encrypted in terms of DNA sequence using substitution cipher. Each character is replaced by DNA triplets using the following encryption key (Table 2).

**Step 2:** The encrypted message in the DNA strand is edged by two 20 bases long forward and reverse PCR primer sequences (Figure 9). Primers are used for amplification of the encrypted segment of the DNA strand.

**Step 3:** The encrypted strands are hidden within fragmented and denatured human genomic DNA. This is the first layer of steganography. Random combination of genomic DNA from different organisms can also be used to increase the complexity of the background. The intended recipient will not be affected by the vast mixtures of genomic DNA as the pair of primer sequences are known to them.

**Table 2:** *Encryption key used in [20].*

| Character | DNA Triplet | Character | DNA Triplet | Character | DNA Triplet | Character | DNA Triplet |
|---|---|---|---|---|---|---|---|
| A | CGA | K | AAG | U | CTG | U | ACT |
| B | CCA | L | TGC | V | CCT | 1 | ACC |
| C | GTT | M | TCC | W | CCG | 2 | TAG |
| D | TTG | N | TCT | X | CTA | 3 | GCA |
| E | GGT | O | GGC | Y | AAA | 4 | GAG |
| G | TTT | Q | AAC | | ATA | 5 | AGA |
| H | CGC | R | TCA | , | TCG | 7 | ACA |
| I | ATG | S | ACG | . | GAT | 8 | AGG |
| J | AGT | T | TTC | : | GCT | 9 | GCG |

forward PCR primer
encrypted message
reverse PCR primer
5'
3'

**Figure 9:** *Encrypted DNA strand.*

**Step 4:** The second layer of steganography is the confinement of the sample into DNA microdot. By this step even the medium containing the encrypted message can be obscured from the adversary.

The intended recipients, who know both the encryption key and primer sequences, can only amplify the encrypted DNA sequence, extract the sequence, read it and finally decrypt the secret message.

By this proposed method individual secret messages can be sent to each of several intended recipients by using single or duplicate microdots. Each recipient uses a unique primer pairs for amplification of the particular message which is sent to him/her. The authors claimed that, by this technique encrypted DNA sequences concealed within the genome can be attached using common adhesives to full stops in a printed innocuous letter. It can be also posted through the U.S. Postal Service. The embedded text can be efficiently decoded. Thus, confidentialmessages can be transmitted securely.

## 3.3. Protection of information in living host [21]

In section 3.2 we have gone through a brief discussion on the proposed method of steganography by [20] where the authors recommended that for information storage DNA sequences can be as reliable as a piece of paper. In this section we will refer the research work [21], where the authors encrypted information in synthetic DNA strands and permanently stored the information in the living host securely allowing the organism to grow and multiply. They ensured the protection of encrypted DNA sequences from the adverse circumstances, such as, fatal double strand break of DNA caused by extreme temperature and desiccation or rehydration; presence of DNA nucleases; ultraviolet ray, ionizing radiation; intentional attack by any individual etc. For preservation of information in the organism in such a way that it can be recovered again, in [20] the following steps are executed.

**Step 1:** The suitable host, which can carry the information and survive in unfavorable environment, is identified. The authors selected two bacteria, *Escherichai coli* (E. coli) and *Deinococcus radiodurans* (Deinococcus), whose whole genomes have been completely sequenced. In addition to this, these organisms can resist extreme conditions.

**Step 2:** The information is encrypted in terms of DNA sequence using a specific encryption key where each character is replaced by DNA triplets.

**Step 3:** This is the most significant step of the experiment where a set of fixed length (20 base pairs) unique DNA sequences has to be identified. The selected string of bases should not existin the host organism and ensure all the genomic limitations. Otherwise, unwanted mutations and damage of the encrypted information may occur. The selected set of sequences also acts as terminal domains of the embedded information. The authors selected a set of 25 DNA oligonucleotides which are 20 bases long (Table 3). These sequences include stop codons (TAA, TGA and TAG) which signal the termination of translation into protein. Otherwise, the production of synthetic proteins from the encrypted DNA sequences may cause the destructionof the information as well as the host organism.

**Step 4:** Synthetic double stranded DNA oligonucleotide (Figure 10), 46 base pairs long, is designedwhich has three domains. The first and last 20 base pairs long domains are selected from Table 3 (sequences must be different) and the 6 base pairs long domain in between is a restriction site which can be cleaved by the corresponding restriction endonuclease. The restriction site is for later incorporation of encrypted DNA sequence.

**Step 5:** The synthetic 46 base pairs long double stranded DNA sequence is then cloned in recombinant plasmid. The 20 base pairs long terminal domains (Figure 10) act as the identification marker of the encrypted information, as these sequences are not present in the host genome.

**Step 6:** The secret information is encoded in terms of DNA sequence using encryption key and then incorporated into the cloning vector using the restriction endonuclease.

**Table 3:** *Selected set of unique DNA oligonucleotide [21].*

| AAGGTAGGTAGGTTAGTTAG | AGAGTAGTGAGGATAGTTAG |
| --- | --- |
| AGGTTTGGTGGTATAGTTAG | ATAAGTAGTGGGGTAGTTAG |
| ATAGGAGTGTGTGTAGTTAG | ATAGGGGTATGGATAGTTAG |
| ATATTAGAGGGGGTAGTTAG | ATGGGTGGATTGATAGTTAG |
| GGAGTAGTGTGTATAGTTAG | GGGAATAGAGTGTTAGTTAG |
| GGGAGTATGTAGTTAGTTAG | GGGATGATTGGTTTAGTTAG |
| GGTTAGATGAGTGTAGTTAG | GTATGGGAATGGTTAGTTAG |
| TAAGGGATGTGTGTAGTTAG | TAGAGAGAGTGTGTAGTTAG |
| TAGAGGAGGGATATAGTTAG | TAGAGTGGTGTGTTAGTTAG |
| TAGATGGGAGGTATAGTTAG | TAGATTGGATGGGTAGTTAG |
| TAGGAGAGATGTGTAGTTAG | TAGGGTTGGTAGTTAGTTAG |
| TATAGGGAGGGTATAGTTAG | TATAGGGTAGGGTTAGTTAG |
| TGTGGGATAGTGATAGTTAG | |

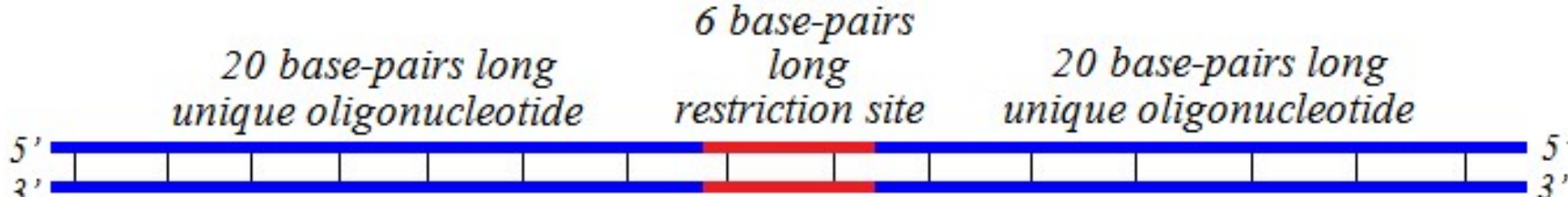

**Figure 10:** *Designed 46 base pairs long DNA sequence.*

**Step 7:** The resultant circular DNA molecule i.e., the vector (Figure 11) is transferred within thehost organism by electroporation. The encoded DNA sequence can amplify with the living cell. The encrypted information is now secure after the incorporation of the recombinant plasmid into the genome of a living host.

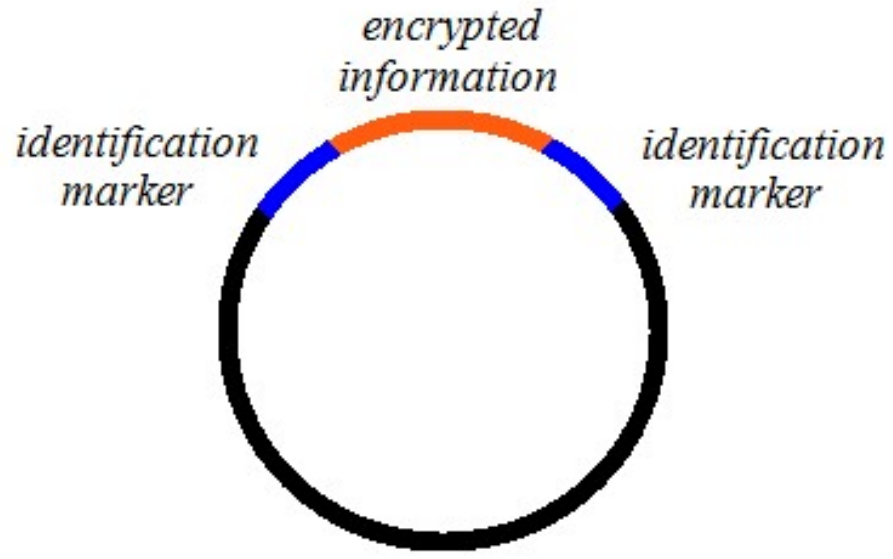

**Figure 11:** *Recombinant plasmid.*

**Step 8:** The encrypted information can be retrieved by the recipient by performing polymerase chain reaction (PCR) in the wet lab. Using the identification marker as the PCR primer sequences, the encrypted DNA segment can be amplified. After reading the extracted amplified sequence, the information can be decrypted using the encryption key.

## 3.4. DNA Cryptography using binary strands [22]

DNA binary strands to perform cryptography was presented in [22]. They showed that if the adversary has same technical potentials as the sender and the receiver of the secret message; even then the proposed cryptosystem works efficiently. They alsoprojected another technique of cryptography based on graphical subtraction of binary gel images.

### 3.4.1. DNA Cryptosystem by generating dummy strands

The representation of digital binary strings in terms of DNA sequences was first developed by [23]. The DNA binary strands encoding digital text with different lengths start and end with predetermined terminator domains termed as *s* and *e* respectively. The coded binary strands are in the form of *s*{0|1}e. Two different types of partially double stranded DNA oligonucleotides with sticky ends are used for representation of 0-DNA bit and 1-DNA bit. Terminator domains also have sticky ends. DNA binary strands are formed by repeated concatenation of the oligonucleotides encoding bits through the complementary sticky ends. Figure 12 shows the DNA binary strands which are the representative of the corresponding digital binary strings.

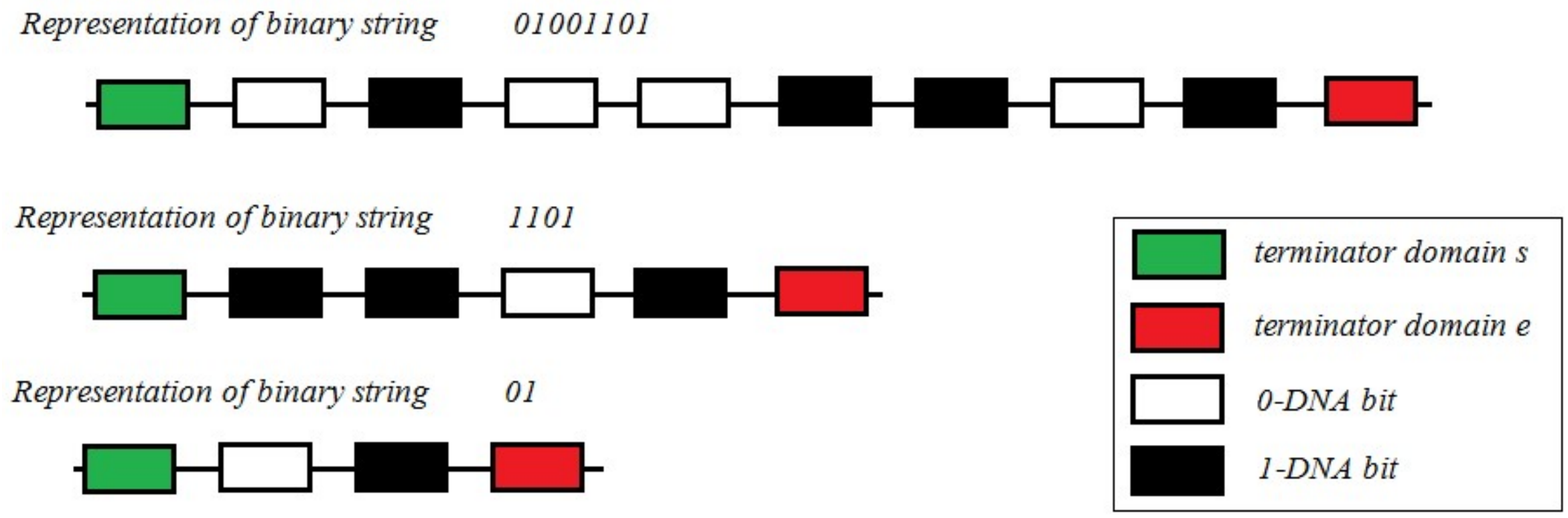

**Figure 12:** *Encoding DNA binary strand.*

The cryptosystem based on DNA steganography proposed by [22] is explained through the following steps.

**Step 1:** The encryption key i.e., the unique identification sequence is shared between the sender and receiver of the secret message through secure communication channel. The key sequence canbe terminator domain of the binary strand.

**Step 2:** The secret message i.e., the digital binary string is encrypted in the form of DNA sequence. The key sequence is ligated to the encrypted strand.

**Step 3:** A certain number of dummy DNA strands are generated which have similar binary format as the encrypted strand. This is because of the fact that the encrypted strand follows a particular linguistic structure (for example, English); but if the dummy strands are generated randomly, then, the adversary may take benefit of this particular dissimilarity between the encrypted strand and the dummy strands.

**Step 4:** The dummy strands and the encrypted strands are mixed in equimolar amounts.

**Step 5:** The resultant solution is sent to the intended receiver through an open communication channel.

**Step 6:** The encrypted message can only be decoded by the receiver who knows the encryption key. Using the key sequence as one of the primers and the corresponding 0-DNA bit or 1-DNA bit as another primer PCR is performed.

**Step 7:** Gel electrophoresis is performed using the amplified sequences. The encrypted strand is extracted from gel and decrypted.

Though it has been assumed that the adversary has the same technical potentials as the sender and the receiver, the possibility to differentiate between the encrypted strand and the dummy strands by the adversary is very low. The only line of attack is to guess the key sequence or extract the encrypted strand coincidentally, which is also very rare. If the *security* of the proposed cryptosystem is $\sigma$, then probability of randomly selection the encrypted strand is $(1-\sigma)$, which is very low.

### 3.4.2. DNA Cryptosystem by graphical subtraction of gel image

Another unconventional technique of steganography was projected by [22], where predetermined key sequence is not used for decryption. The pool of dummy strands is used for decoding the hidden message. Gel electrophoresis has a significant function in this proposed technique. The security of the cryptosystem is based on the graphical subtraction of the gel images by the existing methods of digital image processing. There are three variations of this technique based on the encryption key.

1. **Pool of dummy strands is the key.** The encrypted strands are mixed with equimolar amounts of the dummy strand. Another set of dummy pool is shared in advance as the encryption key. The receiver executes gel electrophoresis twice following the PCR. The first one is with the encoded solution and the second one is using the key solution. By graphical subtraction of the gel images the decryption of the secret message is achieved.

2. **Gel image of the dummy pool is the key.** In this case directly the gel image of the pool of dummy strands is sent to the receiver as the key instead of sending the solution.

3. **Sequence information of the dummy pool is the key.** The relevant information needed to generate the pool of dummy strands is used as encryption key. The dummy sequences, their frequencies and other physical parameters are shared with the receiver so that he/shecan produce the pool and decrypt the hidden message.

The limitation of this proposed method is the resolution of the gel-image. The authors detected 32 bits long bit-chain using this technique. Higher resolution leads the generation of longer decrypted bit-chain. Thus, further research is needed for incorporation of automation in this method to overcome the limitations.

## 3.5. Symmetric-key DNA cryptosystem [24]

DNA cryptosystem based on symmetric key technique was proposed in [24]. They have incorporated DNA microarray technology into DNA cryptography to design a secure cryptosystem which is unaffected even by highly efficient quantum computers. The researchers have exploited the massive parallelism of DNA computationand astonishing information storage capacity of DNA molecules to develop DNA Symmetric-key Cryptosystem (DNASC). Symmetric key encryption uses the same key to encrypt and decrypt secret information. The mechanism of DNASC is described briefly through the following steps.

**Step 1:** Short single stranded DNA oligonucleotides attached on the solid surface of DNA chip i.e., DNA probes are used as the encryption key of this mechanism. The complementary sequences of the encryption key are used as the decryption key. Selective standard hybridization condition can also be included as the decryption key along with the complementary DNA probes.

The probes are selected from a known experiment which has to be kept secret. Here, the researchers have used known microarray data which demonstrates diauxic shift in the yeast i.e., gene transcription spectrum from anaerobic (fermentation) to aerobic (respiration) metabolism [25]. If $\beta$ is the set of encryption key; the subset $\beta_1$ represents the binary digit "1"and the subset $\beta_0$ represents the binary digit "0" depending on the hybridization intensity. It can be concluded that the encryption key, $E_K = \beta_1 \cup \beta_0$.

The decryption key $D_k$ i.e., set $a$ which is complementary to the set $\beta$, is sent to the intended receiver through secure channel.

**Step 2:** The secret message is translated into digital binary matrix which is written into a binary virtual chip using the encryption key. The encrypted chip is sent to the receiver through an insecurechannel.

**Step 3:** The receiver, who knows the set $a$ and standard hybridization conditions, can decrypt the message by the process of hybridization. The hybridization signal is translated into digital binary matrix by the standard methods of signal processing. The coloured spots of the chip represent thebinary digit "1" and the black spots represent the binary digit "0". Thus, secret plaintext can be recovered.

There are two levels of security in this proposed technique. The first level is biological security. DNASC does not use any mathematical computation; it completely depends on the complex biological processes. To break this cryptosystem, the adversary is supposed toget not only the molecular information but also the ratio of each kind of probe on each spot of encrypted chip, which is nearly impossible. The popular sequencing methods, viz. Maxam-Gilbert method and Sanger's method, fail to read the small oligonucleotides on a chip. The researchers have made the encryption chip more complicated by using mixed sequence-specific probes. Thus, effective separation of DNA probes from a mixture of unknown oligonucleotides is almost unachievable.

The second layer of security is computational security. After attempting for many years, if the biological security breaks, then the computational security comes into action. It hides in the coding method of DNASC applied to the DNA microarray [24].

## 3.6. Asymmetric-key DNA cryptosystem and signature method [26]

The asymmetric-key DNA cryptosystem and signature method was introduced in [26]. The authors have designed DNA-PKC which can perform asymmetric key encryption and mimic digital signature method using DNA microarray technology.

In this methodology one set of DNA probes represents public key ($E_k$) and the other setof DNA probes represents the private key ($D_k$). Along with the probes selective standard hybridization condition can also be included as the decryption key. Using the public key everyone can encrypt secret message on a DNA chip and physically transmit it to the receiver. Only the intended recipient who has the private key can decrypt the ciphertext. The set of private key are designed in such a way that it has certain connection with the public key. In this model the probes of private key are complementary to the public key for hybridization purpose. On the DNA chip if the probes get hybridized according to predetermined standard, it will produce signal with variable intensity. If the signal intensity is higher than a predefined threshold value, itrepresents binary digit “1” and the probes with signal intensity lower than the threshold value represents the binary digit “0”. For decryption the hybridization signal is translated back to the original text.

In signature methodology, the private signing key holder can generate a signature and only the public verification key holders can authenticate the signature. Let us assume that a person *A* wants to authenticate the signature generated by *B*. The mechanism of this method is discussed briefly through the following steps.

**Step 1:** A key pair ($V_k$, $S_k$) is randomly chosen by *B* from which the private signing key $S_k$ is keptsecret. The public verification key $V_k$ is amplified and distributed to all the verifiers.

**Step 2:** *B* encrypt the secret message on DNA chip using the secret key $S_k$. The chip carrying theciphertext is sent to the intended receivers through insecure channels.

**Step 3:** Among the intended receivers, *A* also receives the encrypted DNA chip. Using a share of *B*’s verification key $V_k$, *A* allows the hybridization on the DNA chip to authenticate the signature of *B*. If another receiver *C* wants to authenticate *B*’s signature, then he/she will follow the same procedure using his/her share of the verification key which might be different from that of *A*’s. Neither of the recipients can forge the signature.

Similar to the symmetric key DNA cryptosystem described in section 3.5 [24], this technique also has two layers of security. One is biological security and the otheris computational security. Another advantage of the proposed *DNA-PKC* is that the private key cannot be retrieved from the known public key.

## 3.7. Triple stage DNA cryptography [27]

A new algorithm of DNA cryptography based on the concept of Moore machine in automata theory was suggested by [27]. The authors claim that the designed cryptosystem is more dependable, and the security of this technique relies on the three encryptionstages which use secret key, auto generated Moore machine and password.

Automata theory can be defined as the study of abstract self-propelled devices which follow predetermined sequence of operations to solve computational problem automatically. Final state machine (FSM) is a type of automaton that can be defined as a machine with finite number of states and the automaton can be in exactly one state at any given time. Moore machine is a FSM which produces output depending only on the present state. It can be defined by a 6 tuple (Equation 5).

$$M = (Q, \sum, \Delta, \delta, \lambda, q_0) \quad [5]$$

where,

$Q$ is a nonempty finite set of states.
$\sum$ is a nonempty finite set of input symbols.
$\Delta$ is a finite set of output symbols.
$\delta$ is the input transition function, where $\delta: Q \times \sum \rightarrow Q$
$\lambda$ is the output transition function, where $\lambda: Q \rightarrow \Delta$
$q_0$ is the initial state, where $q_0 \in Q$.

The algorithm presented in [27] is briefly discussed by the following steps;

**Step 1***:* The input hidden message is first encrypted by using user generated/ dynamically generated secret key.

**Step 2:** The encrypted data is converted into binary sequence. The binary data is partitioned in $n$ number of blocks of size 256 bits each.

**Step 3:** The XOR operation is performed on the blocks.

**Step 4:** The output of the XOR operation is converted into DNA sequence following the codebook-I (Table 4).

**Step 5:** In this step Moore machine takes the resultant DNA sequence of step 4 as input and produces user password data as output. The transition table (Table 5) for generating user password data is given below.

**Step 6:** The newly created user password data is again converted in DNA sequence following DNA codebook-II (Table 6). The resultant sequence is the final encrypted data which is sent to the intended recipient.

Because of the multiple layers in the encryption methodology, the authors assert that the proposed algorithm is very difficult to break. The added advantage of the algorithm is handling a broad range of plaintext in terms of data size.

**Table 4:** *DNA codebook-I.*

| Binary Value | Code | Binary Value | Code |
|---|---|---|---|
| 0000 | AA | 0101 | CC |
| 1000 | AG | 1001 | CG |
| 0100 | AC | 0001 | CA |
| 1100 | AT | 1101 | CT |
| 0010 | GA | 0011 | TA |
| 1010 | GG | 1111 | TT |
| 0110 | GC | 0111 | TC |
| 1110 | GT | 1011 | TG |

**Table 5:** *State transition table for user password data generation.*

| Current State | Next State | | | | Output |
|---|---|---|---|---|---|
| | Input = *A* | Input = *G* | Input = *T* | Input = *C* | |
| $q_0$ | $q_1$ | $q_7$ | $q_{12}$ | $q_{17}$ | *Null* |
| $q_1$ | $q_2$ | $q_3$ | $q_4$ | $q_5$ | *Null* |
| $q_2$ | $q_1$ | $q_7$ | $q_{12}$ | $q_{17}$ | $W_1$ |
| $q_3$ | $q_1$ | $q_7$ | $q_{12}$ | $q_{17}$ | $W_2$ |
| $q_4$ | $q_1$ | $q_7$ | $q_{12}$ | $q_{17}$ | $W_3$ |
| $q_5$ | $q_1$ | $q_7$ | $q_{12}$ | $q_{17}$ | $W_4$ |
| $q_7$ | $q_8$ | $q_9$ | $q_{10}$ | $q_{11}$ | *Null* |
| $q_8$ | $q_1$ | $q_7$ | $q_{12}$ | $q_{17}$ | $W_5$ |
| $q_9$ | $q_1$ | $q_7$ | $q_{12}$ | $q_{17}$ | $W_6$ |
| $q_{10}$ | $q_1$ | $q_7$ | $q_{12}$ | $q_{17}$ | $W_7$ |
| $q_{11}$ | $q_1$ | $q_7$ | $q_{12}$ | $q_{17}$ | $W_8$ |
| $q_{12}$ | $q_{13}$ | $q_{14}$ | $q_{15}$ | $q_{16}$ | *Null* |
| $q_{13}$ | $q_1$ | $q_7$ | $q_{12}$ | $q_{17}$ | $W_9$ |
| $q_{14}$ | $q_1$ | $q_7$ | $q_{12}$ | $q_{17}$ | $W_{10}$ |
| $q_{15}$ | $q_1$ | $q_7$ | $q_{12}$ | $q_{17}$ | $W_{11}$ |
| $q_{16}$ | $q_1$ | $q_7$ | $q_{12}$ | $q_{17}$ | $W_{12}$ |
| $q_{17}$ | $q_{18}$ | $q_{19}$ | $q_{20}$ | $q_{21}$ | *Null* |
| $q_{18}$ | $q_1$ | $q_7$ | $q_{12}$ | $q_{17}$ | $W_{13}$ |
| $q_{19}$ | $q_1$ | $q_7$ | $q_{12}$ | $q_{17}$ | $W_{14}$ |
| $q_{20}$ | $q_1$ | $q_7$ | $q_{12}$ | $q_{17}$ | $W_{15}$ |
| $q_{21}$ | $q_1$ | $q_7$ | $q_{12}$ | $q_{17}$ | $W_{16}$ |

**Table 6:** *DNA codebook-II.*

| Password Letters | DNA Code | Password Letters | DNA Code |
|---|---|---|---|
| $W_1$ | A | $W_2$ | T |
| $W_3$ | C | $W_4$ | G |
| $W_5$ | AG | $W_6$ | AC |
| $W_7$ | AT | $W_8$ | GC |
| $W_9$ | GT | $W_{10}$ | GA |
| $W_{11}$ | TC | $W_{12}$ | TG |
| $W_{13}$ | TA | $W_{14}$ | CA |
| $W_{15}$ | CT | $W_{16}$ | CG |

### 3.8. DNA linear block codes [28]

Recently, Mondal and Ray [28] used the notion of DNA linear block codes to generate DNA codewords with an aim to transmit data securely. For designing the codewords a new design strategy, DNA-based XOR operation (DNAX), has been applied effectively. In the proposed methodology the information was first encrypted into short single stranded DNA sequences, termed as DNA codewords, and can be decrypted again. The generator matrix and parity check matrix are used for encryption and decryption of data. This codeword design strategy can effectively detect and correct errors in transmission through biological channels.

## 4. Conclusion

DNA cryptography is a promising and rapidly emerging field in data security. The conventional binary data uses two digits '0' and '1' to code information. But for DNA molecules, which are the natural transporter of information, data is encoded by four bases viz. 'A','T', 'G' and 'C'. A few grams of DNA molecules have the capacity to restrain all the data stored in the world. Adleman has explored [29] how the huge parallelism of DNA strands can concurrently attack the different aspects of the toughest combinatorial problem and solve it in polynomial time. DNA cryptography merges the massive parallelism and storage capacity of DNA molecules with the traditional methodologies of cryptography. It also has the capability for detecting and correcting error while transmitting data [28]. At present, big tech giants, such as Microsoft, are taking the initiative to commercialize DNA computers in the near future. Hopefully, in the coming ten to thirty years the virtually unhackable DNA cryptography techniques will be an effective alternative to classical cryptosystem.

Modern biological science is becoming gradually more digitized which is beneficial in cyberbiosecurity. Storage of genome information in electronic databases is the keystone of modern digitized biotechnology. The increased use of computers in cyberbiosecurity and capability to manipulate DNA strands explore elevated risk in deliberate destruction of biological equipment and production of hazardous biological materials. So, with the development of DNA cryptography, we should keep an eye on the possible pitfalls of budding technology.

## Acknowledgement

The first author acknowledges the financial support received as Research Associate fellowship from Council of Scientific & Industrial Research: Human Resource Development Group (CSIR: HRDG), Government of India.